\documentclass[aps,pra,twocolumn,showpacs,superscriptaddress]{revtex4} 
\usepackage{graphicx}
\usepackage{amsmath}
\usepackage{amssymb}
\usepackage{epstopdf}
\usepackage{amsfonts}
\usepackage{dcolumn}
\usepackage{bigints}
\usepackage{xcolor}
\usepackage[normalem]{ulem}
\begin{document}
\title{Non-exponential tunneling
due to mean-field induced swallowtails}
\author{Q. Guan}
\address{Homer L. Dodge Department of Physics and Astronomy,
  The University of Oklahoma,
  440 W. Brooks Street,
  Norman,
Oklahoma 73019, USA}
\address{Center for Quantum Research and Technology,
  The University of Oklahoma,
  440 W. Brooks Street,
  Norman,
Oklahoma 73019, USA}
\author{M.~K.~H. Ome}
\address{Department of Physics and Astronomy, Washington State
University, Pullman, Washington 99164-2814, USA}
\author{T.~M. Bersano}
\address{Department of Physics and Astronomy, Washington State
University, Pullman, Washington 99164-2814, USA}
\author{S. Mossman}
\address{Department of Physics and Astronomy, Washington State
University, Pullman, Washington 99164-2814, USA}
\author{P. Engels}
\address{Department of Physics and Astronomy, Washington State
University, Pullman, Washington 99164-2814, USA}
\author{D. Blume}
\address{Homer L. Dodge Department of Physics and Astronomy,
  The University of Oklahoma,
  440 W. Brooks Street,
  Norman,
Oklahoma 73019, USA}
\address{Center for Quantum Research and Technology,
  The University of Oklahoma,
  440 W. Brooks Street,
  Norman,
Oklahoma 73019, USA}
\date{\today}

\begin{abstract}
Typically, energy levels change without bifurcating in response to a change of 
a
control parameter.
Bifurcations can lead to loops or swallowtails in the energy spectrum.
The simplest quantum Hamiltonian that supports 
swallowtails
is a 
non-linear $2 \times 2$ Hamiltonian with 
non-zero off-diagonal elements and
diagonal elements that depend on the 
population difference of the two states.
This work implements such a  Hamiltonian experimentally using ultracold atoms in
a moving one-dimensional 
optical lattice.
Self-trapping and  
non-exponential tunneling
probabilities, a hallmark signature of band structures that support swallowtails, 
are observed.
The good agreement between theory and experiment validates the optical lattice
system as a
powerful platform to study,
 e.g., 
Josephson junction
physics
and 
superfluidity in ring-shaped geometries.
\end{abstract}
\maketitle

In time-dependent processes, two limiting scenarios
are of particular interest:
the regime where the system Hamiltonian is quenched (i.e.,
changed essentially instantaneously)
and the opposite 
regime where the system Hamiltonian is changed adiabatically
(i.e., so 
slowly
that transitions between different adiabatic
eigenstates are strongly suppressed).
Generally, the adiabatic regime is reached when the ramp 
rate $\alpha$,
with which the control parameter $\gamma$ is changed,
 is sufficiently small compared to the 
rate
that is set by 
the energy gap $\Omega$ ($\Omega$ is 
taken to be real) at the avoided 
crossing of neighboring adiabatic eigenstates.
This is captured by the celebrated ``linear'' 
Landau-Zener formula~\cite{ref_landau,ref_zener}, which gives the tunneling 
probability $r$ between two energy levels,
assuming 
$\gamma$
changes
linearly
with time $t$ 
[$\gamma(t)=\alpha t$, $\alpha  > 0$],
\begin{eqnarray}
\label{eq_linear_lz}
r = \exp \left[ - \pi \Omega^2/ (2 \hbar \alpha) \right].
\end{eqnarray}
According to the Landau-Zener formula, 
adiabaticity (i.e., the $r \rightarrow 0$ limit)
can always be 
approached,
at least in principle, by reducing 
the ramp rate $\alpha$.

\begin{figure}
\includegraphics[width=0.4\textwidth]{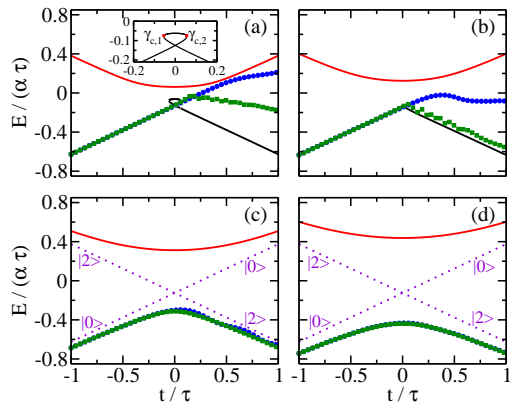}
\caption{
Scaled energy levels
of $\hat{H}_{\text{TS}}$ for $C/(\alpha\tau) = 0.268$
as a function of 
$t/\tau$
for
(a) $C/\Omega= 2.14$,
(b) 1.07,
(c) 0.428,
and
(d) 0.306.
The black and red solid lines show the adiabatic energy levels
[in panels~(c) and (d), the black lines are covered by the symbols].
The blue circles and green squares show the dynamic energy levels
for 
$\hbar/(\alpha\tau^2) = 7.68 \times 10^{-2}$ and $\hbar/(\alpha\tau^2) = 7.68 \times 10^{-3}$, respectively
[in (d), the blue circles are covered by the green squares].
The tunneling probability is appreciable in (a) and (b) and essentially zero in (c)
and (d).  
The inset in (a) shows an enlargement of the swallowtail;
$\gamma_{\text{c}, 1}/(\alpha \tau)$ and $\gamma_{\text{c}, 2}/(\alpha \tau)$ 
correspond to the boundaries of the swallowtail. 
As a reference, the purple dotted lines in (c) and (d)
show the adiabatic energy levels for 
$\Omega=0$;  the energy levels are labeled by their eigenstates.
The parameters used to make the solid lines and blue circles are the same as those used 
in Fig.~\ref{fig2}.
}
\label{fig1}
\end{figure}

The presence of a non-linearity $C$ 
alters the tunneling dynamics qualitatively and
quantitatively~\cite{niu1,zobay,niu2,niu3,smith1,mueller,smith2,morsch1,morsch3,morsch4,witthaut2006,morsch5,morsch2,bloch,gadway,zhang2019}.
Adiabaticity breaks down
for certain parameter
combinations of the non-linear two-state model, i.e., even an infinitely slow ramp
induces non-adiabatic
population transfer between states, and
the tunneling probability is not given by the  ``standard exponential"~\cite{niu2}.
The breakdown of adiabaticity is intimately linked to the 
phenomenon of hysteresis and the existence of
swallowtails in the adiabatic energy levels of the non-linear
two-state model~\cite{mueller,campbell}.
Mapping to a classical Hamiltonian shows that the 
swallowtail structure emerges
when two new fixed points,
one stable and the other unstable, are first supported 
for 
$\gamma=\gamma_{\text{c},1}$; see the inset of Fig.~\ref{fig1}(a)~\cite{niu2,liu2003}. 
As the control parameter $\gamma$
crosses $\gamma_{\text{c},2}$ ($\gamma_{\text{c},2}> \gamma_{\text{c},1}$), 
a stable and an unstable fixed point collide
and annihilate. In this picture,
the associated 
homoclinic
orbit
is responsible for deviations from adiabaticity~\cite{niu2}.
While the non-linear two-state model
captures aspects of a wide range of systems such as the motion of 
small polarons~\cite{small-polaron1,small-polaron2},
Josephson junctions~\cite{silver,mullen,boshier},
helium and other superfluids in annular rings~\cite{campbell,fetter,packard,campbell2,campbell3},
and Bose-Einstein condensates (BECs) in optical 
lattices~\cite{niu5,choi2003,koller2016,watanabe2016},
non-exponential tunneling
originating from swallowtails
has not yet been
demonstrated experimentally.

Using ultracold $^{87}$Rb atoms 
in a moving one-dimensional 
optical lattice,
the present joint experiment-theory study
investigates two-state dynamics in the presence of swallowtails.
The main results are:
First, 
 a breakdown of adiabaticity  is observed.
The experimental data are reproduced by mean-field Gross-Pitaevskii (GP)
equation simulations 
and interpreted in terms of self-trapping due to mean-field
interactions.
Second, 
non-exponential tunneling probabilities are observed
for parameter combinations for which the adiabatic band structure supports 
swallowtails.
Third, intriguing internal dynamics
are 
revealed 
despite the fact that the initial BEC has a momentum distribution that is narrow 
compared to the size of the Brillouin zone.

Consider the time-dependent Schr\"odinger equation~\cite{niu1}
$\imath \hbar
\partial _t{{\vec{b}}(t)}
= \hat{{H}}_{\text{TS}}
\vec{b}(t)$,
where the non-linear $2 \times 2$ Hamiltonian
$\hat{H}_{\text{TS}}$ is given by
\begin{eqnarray}
\label{eq_hnl2}
\hat{H}_{\text{TS}}
= \frac{1}{2}
\left(
\begin{array}{cc}
\gamma(t)- C \Delta b(t)& \Omega\\
\Omega & -\gamma(t) + C \Delta b(t)
\end{array}
\right)
\end{eqnarray}
and the state vector
$\vec{b}(t)$ by $\vec{b}(t)=(b_0(t),b_2(t))^T$.
The subscripts  ``0" and  ``2" are used 
since our experimental realization connects 
two sites of a momentum lattice, one with
momentum zero
and one with momentum $2 \hbar k_L$~\cite{niu2}, where $k_L$ denotes the
lattice wave vector  (see below for details). 
In Eq.~(\ref{eq_hnl2}), $\Delta b(t)$ denotes the population 
imbalance, $\Delta b(t) = |b_0(t)|^2-|b_2(t)|^2$ with normalization
$|b_0(t)|^2+|b_2(t)|^2=1$.
For $-\tau \le t \le \tau$, the control parameter $\gamma(t)$ changes linearly from
$\gamma=-\alpha \tau$ to $\alpha \tau$.

We first consider the case of vanishing non-linearity ($C=0$).
Starting in state $\vec{b}(t)=(1,0)^T \equiv |0 \rangle$ at $t=-\tau$,
the probabilities to be in states $|0 \rangle$ and
$|2 \rangle \equiv (0,1)^T $ at time $\tau$
are in the $\tau \rightarrow \infty$ limit given
by $r$ and $1-r$, respectively.
In practice, $\tau$ is  finite and
the finite time window defines the
 ``dynamic" energy scale $U_d$, $U_d=\hbar/\tau$~\cite{SM}. In addition, 
 $\hat{H}_{\text{TS}}$ is characterized by the ``static" energy scale $U_s$, $U_s=\alpha \tau$,
 and the coupling strength $\Omega$. 
 For Eq.~(\ref{eq_linear_lz}) providing---``on  average"---a
 reliable description of the state populations at the end of the ramp,
 we need $\Omega/(\alpha \tau) \ll 1$;
we use the term ``on average" since the finite time window introduces oscillations around the 
smooth exponential given in Eq.~(\ref{eq_linear_lz})~\cite{SM}.

We now turn to the non-linear two-state model.
The solid lines in Fig.~\ref{fig1} show the adiabatic energy levels of 
$\hat{H}_{\text{TS}}$ for $C/(\alpha\tau) = 0.268$
as a function of $t/\tau$  for
four different $C/ \Omega$.
The band structure displays a swallowtail centered at 
$t=0$ 
for $C / \Omega > 1$ but not
for $C / \Omega < 1$.
The blue circles and green squares show the ``dynamic" energy levels 
of $\hat{H}_{\text{TS}}$~\cite{niu1} for 
two different ramp rates, parametrized by the scale ratio
$\hbar/(\alpha\tau^2)$~\cite{SM}.
For a given parameter combination,
the dynamic energy level is obtained by
calculating the energy expectation value at each time,
using the  lower
adiabatic eigenstate of $\hat{H}_{\text{TS}}$ for $\gamma = -\alpha \tau$ as initial state~\cite{niu1}.
In Figs.~\ref{fig1}(c) and \ref{fig1}(d),
the dynamic energy levels depend rather weakly on 
the 
ramp rate
and agree well with the 
lower adiabatic energy levels.
In this case, the probability to tunnel to the upper adiabatic
energy level during the ramp is very close to zero.
In Figs.~\ref{fig1}(a) and \ref{fig1}(b),
in contrast,
the dynamic energy levels depend on 
$\hbar/(\alpha\tau^2)$ 
and deviate, even for the 
smaller 
$\hbar/(\alpha\tau^2)$ considered
(this corresponds, for fixed $\alpha \tau$, to a slower ramp~\cite{SM}),
from the lower adiabatic energy level.
Deviations persist even for infinitely slow ramp 
rates~\cite{niu1},
i.e., the probability to tunnel to the upper adiabatic
energy level during the ramp is non-zero.

\begin{figure}
\vspace*{0.0in}
\hspace*{0.0in}
\includegraphics[width=0.43\textwidth]{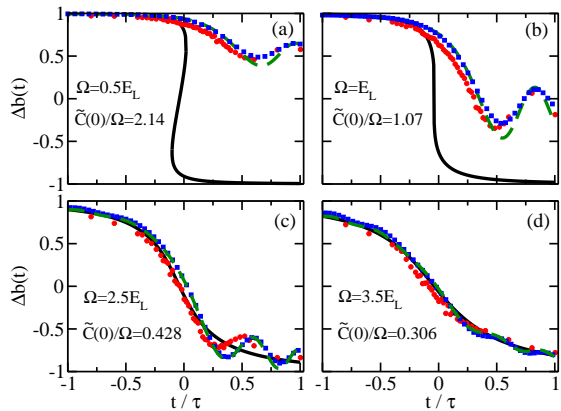}
\vspace*{-0.1in}
\caption{
Experiment-theory comparison of population imbalance
$\Delta b(t)$ in response to a linear ramp with $\alpha= h\times 9$~kHz/ms
corresponding to
$\hbar/(\alpha\tau^2) = 7.68 \times 10^{-2}$]
as a function of 
$t/\tau$ 
for various $\tilde{C}(0)/\Omega$.
The experimental results (red circles) are for $N=3.1\times 10^5$,
$\omega_{x,y,z}=2 \pi \times (147,160,29.8)$~Hz,
and
$E_{L} = h\times 1.08$~kHz.
The blue squares and green dashed lines are
obtained from 
$\hat{H}_{\text{GP}}$  
and 
$\hat{H}_{\text{TS,t}}$, respectively
(analyzing the dynamic states).
In both cases,
the initial state is prepared in an axially 
symmetric trap with $\omega_{\rho}=(\omega_x+\omega_y)/2$.
The mean-field energy $\tilde{C}(t)$
is equal to $1.27$~$E_L$ and $1.07$~$E_L$  for  $t = -\tau$ and $t=0$, respectively.
For comparison, the black solid lines show 
$\Delta b(t)$
for the lower adiabatic eigenstate of $\hat{H}_{\text{TS,t}}$.
}
\label{fig2}
\end{figure}

This work realizes the non-linear 
Landau-Zener model experimentally
by
preparing 
 a single-component BEC consisting of $N$
$^{87}$Rb atoms
of mass $m$
in the 
$|F,m_F \rangle = |1, -1 \rangle$
hyperfine state in an optical dipole trap
and by then adiabatically loading the BEC
into 
a one-dimensional optical lattice 
$V_{\text{lat}}(z)$~\cite{latticeRMP,peik1997,engels,guan2020}.
The optical lattice
is created by two 1064~nm beams [with wave vectors
$\vec{k}_1$ and $\vec{k}_2$, $|\vec{k}_1|=|\vec{k}_2|$,
and angular frequencies $\omega_1(t)$ and $\omega_2(t)$] that cross at an angle of
$\approx \pi/2$,
$V_{\text{lat}}(z, t) = 2 \Omega \cos^2 [ k_L z - \phi(t)/2]$;
$\Omega$
denotes the effective coupling strength, $k_L \approx |\vec{k}_1|/\sqrt{2}$, 
$\phi(t)=[\omega_1(t)-\omega_2(t)]t$
with $\phi(t)=0$ for $t< -\tau$, 
and
$\delta_L(t)=\hbar \partial_t \phi(t)$. 
At $t=-\tau$, the optical dipole trap is turned off and the BEC,
which has an average momentum close to zero, sits
at the  
``bottom"
of the first Brillouin zone; 
this corresponds to a good approximation to state $|0 \rangle$.
In our first set of experiments, $\delta_L(t)$ is---for $t > -\tau$---increased 
linearly 
from $0$ with ramp 
rate
$\alpha = h\times9$~kHz/ms.
The time sequence is designed such that $\delta_L(0)$ is equal to $4$~$E_L$
and 
$\delta_L(\tau)$
is equal to $8$~$E_L$,
i.e., such that the edge of the first Brillouin zone and the middle of the second Brillouin zone 
are reached when $t=0$ and  
$t=\tau$, respectively
[here, $E_L=\hbar^2 k_L^2/(2m)=h \times 1.08$~kHz].
In each repetition of the experiment, the
ramp is stopped at various $t$ and the occupations of the 
components centered at vanishing momentum along the $z$-direction
(state $|0 \rangle$) and centered
at momentum $2 \hbar k_L$ 
(state $|2 \rangle$) are measured after 16.5~ms time of flight, counted from the end of the ramp.
During the time-of-flight expansion, the two momentum components separate fully in 
real space.
The red circles in Fig.~\ref{fig2} show the 
experimentally determined population imbalance  $\Delta b(t)$.
It can be seen that the BEC occupies, for
$t / \tau \gtrsim 0$,
primarily state $|2 \rangle$
when $\Omega$ is ``large"
and primarily state $|0 \rangle$ 
when 
$\Omega$ is ``small".

The lattice system is described by the time-dependent
GP equation with Hamiltonian $\hat{H}_{\text{GP}}$~\cite{latticeRMP},
\begin{eqnarray}
\label{eq_hgp}
\hat{H}_{\text{GP}}=
\hat{\vec{p}}^2/(2m) 
+ V_{\text{lat}}(z, t)  
+g (N-1) |\Psi(\vec{r},t)|^2.
\end{eqnarray}
Here, $g$ is equal to $4 \pi \hbar^2 a_s/m$
and the
mean-field orbital
$\Psi(\vec{r},t)$ is 
normalized according to  $\int |\Psi(\vec{r},t)|^2 d \vec{r} =1$.
For the $|1,-1\rangle$ state of $^{87}$Rb, the $s$-wave scattering 
length $a_s$ is equal to 
$100.4$~$a_{\text{bohr}}$~\cite{scattering_length}.
Following the experimental protocol,
the blue squares in Fig.~\ref{fig2} show our
GP mean-field simulation results.
The good agreement with the experimental data,
including the reproduction of the
oscillatory behavior of the population imbalance for $t \gtrsim 0$
and the small deviations of the population imbalance from $1$ for
large $\Omega$
near
$t \approx -\tau$, 
indicates that the mean-field framework captures the dynamics 
quite accurately.

To bring out the two-state nature of the lattice system,
we write~\cite{niu1,zobay,guan2020} 
$\Psi(\vec{r},t)= \psi_0(\vec{r},t)+\psi_2(\vec{r},t) \exp( 2 \imath k_L z)$,
i.e., we assume that the populations of the $n\hbar k_L$ momentum components
with $n=-2, \pm 4 ,\pm 6,\cdots$
are small~\cite{guan2020}.
Inserting the ansatz into
the non-linear time-dependent GP equation, 
the Supplemental Material~\cite{SM} develops a semi-analytical
framework that  yields a spatially-independent two-state Hamiltonian $\hat{H}_{\text{TS,t}}$.
The Hamiltonian $\hat{H}_{\text{TS,t}}$ is identical to $\hat{H}_{\text{TS}}$ provided the mapping
$\gamma(t) \rightarrow -4E_L+\delta_L(t)$
and
$C \rightarrow \tilde{C}(t)$ is applied.
The time-dependent mean-field energy
$\tilde{C}(t)$,
$\tilde{C}(t)=g (N-1) \bar{n}(t)$, accounts for the fact that the 
BEC expands during the ramp, thereby resulting in a 
decrease of the mean density $\bar{n}(t)$
with increasing $\gamma(t)$.
Since we are interested in non-linear effects, the decrease of the mean-field energy during the ramp places a
constraint on $\alpha$ for a given
$\tilde{C}(-\tau)/\Omega$. 
Figure~S1 in the Supplemental Material~\cite{SM}
shows the adiabatic and dynamic
energy levels of the Hamiltonian $\hat{H}_{\text{TS,t}}$ for the experimental parameters 
used in Fig.~\ref{fig2}. 
Comparison with Fig.~\ref{fig1}
shows that the adiabatic and dynamic energy levels 
supported by $\hat{H}_{\text{TS}}$
and $\hat{H}_{\text{TS,t}}$ agree 
quite well.

\begin{figure}
\includegraphics[height=6.5cm, width=0.4\textwidth, keepaspectratio]{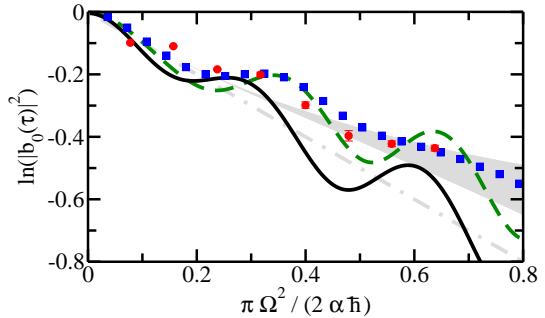}
\vspace*{-0.5in}
\caption{
Experiment-theory comparison of tunneling probability
$|b_0(t)|^2$ at $t=\tau$ in response to linear ramps with varying $\alpha$
for $\tilde{C}(-\tau)/\Omega=2.75$
as a function of $\pi \Omega^2 / (2 \hbar \alpha)$.
The experimental results (red circles;
the error bars  show the standard deviation from three
independent runs) are for $N=2.3\times 10^5$,
$\omega_{x,y,z}=2 \pi \times (193,218,29.8)$~Hz,
$E_{L} = h\times 1.08$~kHz,
and
$\Omega=0.52$~$E_L$.
The blue squares and green dashed lines are
obtained for 
$\hat{H}_{\text{GP}}$  
and 
$\hat{H}_{\text{TS,t}}$, respectively
[analyzing the dynamic states; the initial state is prepared in an axially 
symmetric trap with $\omega_{\rho}=(\omega_x+\omega_y)/2$].
Both data sets
follow the non-exponential trend of the grey-shaded region, which 
shows Eq.~(S11) for $C/\Omega$ values ranging from
$\tilde{C}(-\tau)/\Omega$
to $\tilde{C}(0)/\Omega$~\cite{SM}. 
The black solid line,
which oscillates around the linear Landau-Zener formula 
[grey dash-dotted line; Eq.~(\ref{eq_linear_lz})],
 shows the tunneling probability 
for 
$\hat{H}_{\text{TS}}$ with $C=0$.
The experimental data are better described 
by the non-exponential grey-shaded family of curves than by the linear Landau-Zener formula. 
}
\label{fig3}
\end{figure}    

\begin{figure}
\includegraphics[width=0.48\textwidth]{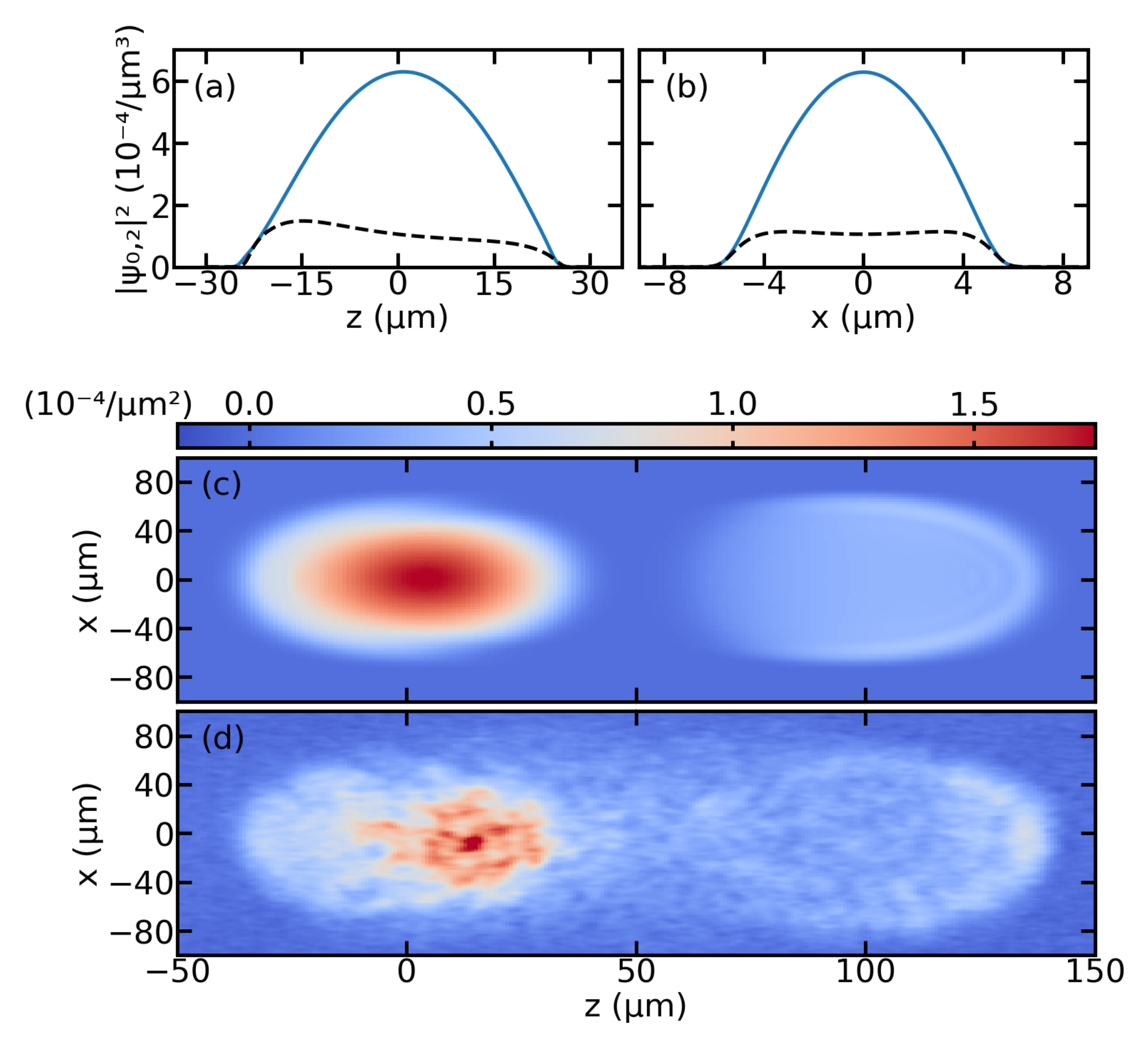}
\caption{Theoretical GP and experimental densities for the
ramp 
ending at $t=-\tau+0.6$~ms 
[$t/\tau=0.25$ in Fig.~\ref{fig2}(b)].
(a) and (b) show density cuts 
before time-of-flight expansion
for $\rho=0$ and $y=z=0$, respectively.
The blue solid and black dashed lines are for states $\psi_0(\vec{r},t)$ and $\psi_2(\vec{r},t)$, respectively.
(c) and (d) show, respectively, theoretical and experimental 
integrated densities $n(x,z,t)$,
$n(x, z, t)=\int_{-\infty}^{\infty} |\Psi(\vec{r}, t)|^2dy$,
after
$16.5$~ms time-of-flight expansion.
}
\label{fig4}
\end{figure}

Black solid and green dashed lines 
in Fig.~\ref{fig2} show the decomposition of the states corresponding to, 
respectively, the lower adiabatic and lower dynamic energy levels
supported by $\hat{H}_{\text{TS,t}}$. 
It can be seen that the green dashed lines agree reasonably well
with the experimental and 
GP results; this
confirms the applicability of the non-linear two-state Hamiltonian to the lattice system.
Moreover, it can be seen that the decomposition of the states corresponding to
the adiabatic and dynamic energy levels agree for the largest $\Omega$ value considered
[Fig.~\ref{fig2}(d)]
 but differ 
for the other $\Omega$ values. 
This shows that 
the system dynamics 
are,
for fixed ramp 
rate
$\alpha$, adiabatic for the largest $\Omega$ considered
in Fig.~\ref{fig2}
but not for the other $\Omega$ values.
In Fig.~\ref{fig2}(c),
the experimental data and populations extracted from the dynamic energy level oscillate around 
the populations extracted from the adiabatic energy level~\cite{Han2015}. 
In Figs.~\ref{fig2}(a) and \ref{fig2}(b), 
the experimental data and populations extracted from the dynamic energy level 
oscillate as well
for $t \gtrsim 0$;
however, the oscillations are not centered
around 
the populations extracted from the adiabatic energy level but instead lie notably above.
Our theory analysis shows that the enhanced tunneling probability (enhanced probability to
remain in state $|0\rangle$) is due to self-trapping, a phenomenon inherently
linked to the presence of swallowtails~\cite{niu5}.

While the inhibition of transitions to state $|2 \rangle$ due to
non-linear interactions has been previously observed in an optical lattice system similar to
ours~\cite{gadway} as well as in coupled double-well type set-ups~\cite{bloch,albiez,steinhauer} 
and annular rings~\cite{campbell},
we now show---for the first time in this context---evidence for non-exponential tunneling.
Red circles in Fig.~\ref{fig3} show the experimentally measured population of state $|0 \rangle$ 
for $t=\tau$ 
and
$\tilde{C}(-\tau)/\Omega = 2.75$; this ratio is a bit larger than
that used in Fig.~\ref{fig2}(a).
It can be seen that the experimental data,
which are obtained by 
varying the ramp 
rate
$\alpha$ (and correspondingly $\tau$ such that $\alpha \tau$
is equal to $4$~$E_L$),
display an overall decrease with increasing $\pi\Omega^2/(2\hbar\alpha)$.
 The experimental data are quite well reproduced by our GP 
 simulations (blue squares).
 The decomposition of the state corresponding
 to the lower dynamic energy level of $\hat{H}_{\text{TS,t}}$ (green dashed line)
 yields notably larger oscillations but displays the same overall trend. 
  In the $\tau \rightarrow \infty$ limit,
 the tunneling probability of the non-linear two-state model $\hat{H}_{\text{TS}}$
 varies  non-exponentially with
 $\pi \Omega^2/(2 \hbar  \alpha)$~\cite{SM}.
 The grey-shaded region shows the results for  $C/\Omega$ values
 between $\tilde{C}(-\tau)/\Omega=2.75$ (upper bound)
 and $\tilde{C}(0)/\Omega$ 
(lower bound; this value
 varies with the ramp 
 rate), respectively.
 The experimental and 
 GP 
 data 
 exhibit small oscillations around 
 the grey region,
 which can be viewed as  a ``smoothed" version of the green-dashed line. 
 For comparison, 
 the black solid line shows the results for the non-interacting two-state model.
Due to the finite time window,
 the black solid line oscillates around the ``linear Landau-Zener" formula 
[Eq.~(\ref{eq_linear_lz}), grey dash-dotted line].
  A key observation of our work is that
 the 
 experimental data are much better described by the non-exponential 
 grey-shaded region than
 the linear Landau-Zener formula.
  Figure~\ref{fig3} provides the first experimental verification of non-exponential tunneling dynamics,
 driven by swallowtails.

Figure~\ref{fig3} 
also
shows that the oscillation amplitude of $\ln[|b_0(t)|^2]$
 is smaller for the experimental and 
 GP 
 data than for the 
 finite-$\tau$ two-state model data. We attribute this 
 to intricate internal dynamics, which 
 are
 not accounted for by the
 two-state models $\hat{H}_{\text{TS}}$ and $\hat{H}_{\text{TS,t}}$.
 To illustrate the internal dynamics, Figs.~\ref{fig4}(a) and \ref{fig4}(b)
 show 
 GP
 densities  for
the ramp stopped at $t/\tau=0.25$ in Fig.~\ref{fig2}(b)
(no time-of-flight expansion).
  The density cuts for the finite momentum component 
 deviate from a simple Thomas-Fermi profile;
in particular, the density along $z$ for $\rho=0$ is deformed, exhibiting a maximum at negative $z$,
 and the density along $x$ for $y=z=0$ exhibits a double peak structure
 [black dashed lines in
 Figs.~\ref{fig4}(a) and \ref{fig4}(b), respectively].
 These density deformations
 develop during the ramp and are attributed to
 the interplay between the on-site and off-site mean-field interactions
 (see Supplemental Material~\cite{SM}).
 
 Figures~\ref{fig4}(c) and \ref{fig4}(d) show
  GP
  and experimentally measured integrated
 densities 
after $16.5$~ms  time-of-flight expansion for the same ramp as considered in Figs.~\ref{fig4}(a) and \ref{fig4}(b).
 The overall agreement between theory and experiment is excellent. The zero-momentum component 
 (centered around $z=0$) has its maximum at positive
 $z$ while the finite-momentum component (centered around $z \approx 100$~$\mu$m)
displays an enhanced density that is located on a  half-ring on the right edge of the cloud.
During the time-of-flight expansion,
 the  finite-momentum component moves relative to the 
zero-momentum component: To reduce mean-field interactions,
the finite-momentum component accumulates  
density first at the left edge of the
cloud and later at the right edge of the cloud.
 The theory data indicate that the relative motion
 of the two clouds generates low energy excitations [wave-like density pattern in Fig.~\ref{fig4}(c)];
although not clearly resolved, faint indications of these patterns are visible in the experimental images.
 
Quantum tunneling is ubiquitous in physics: it plays a central role in high-energy,
nuclear, atomic, and condensed matter physics as well as in chemistry, biology, and engineering.
Modern physics courses introduce students to quantum tunneling and exponentially
decaying tunneling probabilities. The full quantum treatment, however, 
shows that quantum tunneling is much richer, necessitating deviations from the exponential
decay in both the short- and long-time regimes~\cite{fonda,greenland}. 
Indeed, deviations from exponential decay were observed
 in the short-time regime
in a pioneering experiment with cold atoms loaded into 
an accelerated optical lattice~\cite{raizen}. 
The deviations from purely exponential tunneling probabilities observed in this work 
are fundamentally different; they have
their origin in the non-linearity of the interactions. 
Non-linearities also play a fundamental role in the tunneling of a BEC 
out of an external trap into the continuum~\cite{potnis2017,zhao2017}. In that case, however, 
the non-linear Landau-Zener model cannot be applied.
Our work is also fundamentally different from the non-exponential decay 
analyzed theoretically in Floquet-Bloch bands~\cite{weld}, where the 
emphasis lies on short-time deviations and oscillations
due to a finite energy window 
and not due to non-linear mean-field interactions.

{\em{Acknowledgement:}}
Support by the National Science Foundation through
grant numbers
PHY-1806259 (QG and DB) and 
PHY-1607495/PHY-1912540 (MKHO, TMB, SM, and PE) are
gratefully acknowledged.
This work used
the OU
Supercomputing Center for Education and Research
(OSCER) at the University of Oklahoma (OU).

\pagebreak

\begin{center}

\textbf{Supplemental Material: Non-exponential tunnelingSCastin
due to mean-field induced swallowtails}

\end{center}

\setcounter{equation}{0}
\setcounter{figure}{0}
\setcounter{table}{0}
\setcounter{page}{1}
\makeatletter

\renewcommand{\theequation}{S\arabic{equation}}
\renewcommand{\thefigure}{S\arabic{figure}}
\renewcommand{\thetable}{S\arabic{table}}
\renewcommand{\bibnumfmt}[1]{[S#1]}
\renewcommand{\citenumfont}[1]{S#1}

\subsection{Selected properties of $\hat{H}_{\text{TS}}$}
This section discusses  selected properties of the two-state Hamiltonian
$\hat{H}_{\text{TS}}$.
We first discuss why the validity of the linear Landau-Zener formula [Eq.~(1) of the main text] 
requires $\Omega/(\alpha \tau) \ll 1$.
To see this, we consider the adiabatic eigenenergies $\lambda_{\pm}(t)$ of $\hat{H}_{\text{TS}}$,
\begin{eqnarray}
\lambda_{\pm}(t) = \pm \frac{1}{2} \sqrt{ \Omega^2 + \left[ \gamma(t) - C \Delta b(t) \right]^2},
\end{eqnarray}
where $\Delta b(t)$ implicitly depends on $\Omega$, $\gamma(t)$, and $C$.
For $C=0$, the adiabatic eigenenergies can be rewritten as
 \begin{eqnarray}
 \label{eq_sm_eigen2}
\frac{\lambda_{\pm}(t)}{\alpha \tau} = \pm \frac{1}{2} \sqrt{ \left(\frac{t}{\tau} \right)^2+ \left(\frac{\Omega}{\alpha \tau} \right)^2} .
\end{eqnarray}
The derivation of the linear Landau-Zener formula assumes that the initial state
$\vec{b}(-\tau)=(1,0)^T$ is, to a very good approximation, 
equal to the lowest eigenstate of $\hat{H}_{\text{TS}}$ for $C=\Omega=0$ and $t=-\tau$;
this eigenstate has an energy of $- \alpha \tau/2$.
Looking at Eq.~(\ref{eq_sm_eigen2}), the condition on the initial state can be 
expressed as $\lambda_-(-\tau)$ being
approximately equal to $-\alpha \tau /2$ for $t=-\tau$. This condition translates to $[\Omega/(\alpha \tau)]^2 \ll 1$.

Since  $\alpha \tau$ emerges as the reference energy scale
when analyzing the adiabatic eigenenergies, we refer to it as  ``static" energy scale $U_s$. The 
natural time scale associated with 
$U_s$ is given by $T_s=\hbar/U_s$.
To obtain the dimensionless ramp 
rate
$\tilde{\alpha}$, we need to divide
$\alpha$ by $U_s/T_s$. This yields $\tilde{\alpha}=\hbar/(\alpha \tau^2)$. Somewhat counterintuitively, 
the dimensionless ramp 
rate
$\tilde{\alpha}$ is inversely proportional to $\alpha$.
When $\alpha \tau$ is fixed
(this is the case in the experiments discussed in Figs. 2-4 of the main text), 
it is most natural to think about the ramp in terms of the dimensionless ramp 
rate
$\tilde{\alpha}$.
In Fig. 3 of the main text, e.g., the ramp
rate $\alpha$ changes from $37.2$~$E_L$/ms (left most red circle)
to
$4.52$~$E_L$/ms (right most red circle);
these values correspond to $\tilde{\alpha}=0.343$ and $0.0416$, respectively.

Alternatively, we may choose $\tau$ as our natural time unit.
In this alternative set of units, the energy unit is given by $U_d$, $U_d=\hbar/\tau$.
We refer to 
$U_d$
as  ``dynamic" energy scale, since it emerges by defining the
time unit through $\tau$. In these alternative units, the dimensionless ramp 
rate
is
given by $\alpha \tau^2 / \hbar$ (i.e., by $\tilde{\alpha}^{-1}$).
The dimensionless ramp 
rate
$\tilde{\alpha}$ is equal to the scale ratio  $U_d/U_s$.
The
$\tilde{\alpha}\ll 1$ regime corresponds to $U_s \gg U_d$; this is the adiabatic regime.

\subsection{Derivation of two-state  Hamiltonian}
\label{two_state}

Starting with the time-dependent 
GP 
equation, 
this section derives the spatially independent non-linear two-state
Hamiltonian $\hat{H}_{\text{TS,t}}$.
The states $\psi_0(\vec{r},t)$
and $\psi_2(\vec{r},t)$, which are introduced in the main text,
are assumed to be localized in the vicinity
of the momenta $\hbar k_z=0$
and  $\hbar k_z = 2 \hbar k_L$, 
respectively,
and normalized 
such that
$\sum_{j=0,2} \int |\psi_j(\vec{r},t)|^2 d\vec{r}=1$.
As in Ref.~\cite{Sguan2020}, we assume that the widths of the momentum distributions 
associated with the
states
${\psi}_0({\vec{r}},t)$ and
${\psi}_2({\vec{r}},t)$ 
are narrow compared to $2 \hbar k_L$.
For the initial states used in Figs.~2 and 3 of the
main text, the 
full-width-half-maxima
of the momentum distributions along the $z$-direction are
$\hbar\times 0.149$~$\mu$m$^{-1}$ and $\hbar \times 0.141$~$\mu$m$^{-1}$, respectively; for comparison, $2\hbar k_L$ is equal to
$\hbar\times 8.62$~$\mu$m$^{-1}$ (i.e., roughly $60$ times larger).
As a result, we find the following approximate spatially- and time-dependent 
mean-field Hamiltonian~\cite{Sguan2020}:
\begin{widetext}
\begin{eqnarray}
\label{coupled_GP_5}
\hat{{{H}}}= 
I_2 \hat{H}_0 +\hat{H}_1 + \frac{1}{2}
\begin{pmatrix}
   \gamma(t)-g (N-1) (|{\psi}_0({\vec{r}},t)|^2-|{\psi}_2({\vec{r}},t)|^2) &  \Omega\\
    \Omega & -\gamma(t)+g (N-1) (|{\psi}_0({\vec{r}},t)|^2-|{\psi}_2({\vec{r}},t)|^2)\\
\end{pmatrix} ,
\end{eqnarray}
where
\begin{eqnarray}
I_2 = 
\begin{pmatrix}
1 &0\\
0 & 1
\end{pmatrix},
\end{eqnarray}
\begin{eqnarray}
\label{H_0}
\hat{H}_0=\frac{\hat{\vec{p}}^2}{2m}+\frac{3 g (N-1)}{2} \left(|{\psi}_0({\vec{r}},t)|^2 + |{\psi}_2({\vec{r}},t)|^2\right) + \frac{4E_L -\delta_L(t)}{2},
\end{eqnarray}
\begin{eqnarray}
\label{H_1}
\hat{H}_1 = 
\frac{2\hbar k_{L}}{m}
\begin{pmatrix}
0 &0\\
0 & \hat{p}_z
\end{pmatrix},
\end{eqnarray}
and
\begin{eqnarray}
\label{gamma}
\gamma(t) = -4E_L + \delta_L(t).
\end{eqnarray}
\end{widetext}

We now make the ansatz  
 that the spatial orbitals $\psi_j(\vec{r},t)$ for the $j=0$ and $j=2$ components are identical
and that
the occupations of the two components are parametrized by $|b_j(t)|^2$,
\begin{eqnarray}
\psi_j(\vec{r}, t) = b_j(t) \varphi_{\text{TF}}(\vec{r}, t),
\end{eqnarray}
where the normalizations read $\int |\varphi_{\text{TF}}(\vec{r},t)|^2 d \vec{r}=1$ and $|b_0(t)|^2 + |b_2(t)|^2=1$.
At $t=-\tau$, the spatial orbital $\varphi_{\text{TF}}(\vec{r},t)$
is equal to the Thomas-Fermi orbital for a harmonically trapped
$N$-particle BEC. 
We then assume that the component densities maintain their Thomas-Fermi
shape during the ramp. Specifically, we assume that $\varphi_{\text{TF}}(\vec{r},t)$
expands during the ramp in the same manner as a single-component $N$-atom BEC.
This implies that
the time evolution of $\varphi_{\text{TF}}(\vec{r},t)$ is governed by the self-similar solutions derived
by Castin and Dum~\cite{SCastin}.
The adapted formulation neglects the relative motion along the
$z$-direction of the two components with respect to each other.
Moreover, the formulation does not allow for
deviations from the Thomas-Fermi density (structure formation). Correspondingly,
the description should work best for fast ramps and deteriorate for slower ramps.

Integrating over the spatial degrees of freedom, we find the
Hamiltonian $\hat{H}_{\text{TS,t}}$,
 \begin{eqnarray}
 \label{eq_twostate_appendix}
\hat{H}_{\text{TS,t}}= \nonumber \\
\frac{1}{2}
  \left( 
 \begin{array}{cc}
 \gamma(t)-\tilde{C}(t)\Delta b(t) & \Omega\\
\Omega  & 
-\gamma(t)+\tilde{C}(t)\Delta b(t)
 \end{array}
 \right) ,
  \end{eqnarray}
  where
    \begin{eqnarray}
  \label{C_t}
  \tilde{C}(t) = g (N-1) \int |\varphi_{\text{TF}}(\vec{r}, t)|^4d\vec{r}.
  \end{eqnarray}
  Here $\Delta b(t)$ is, as in the main text, equal to $|b_0(t)|^2-| b_2(t)|^2$.
  In going from Eq.~(\ref{coupled_GP_5}) to Eq.~(\ref{eq_twostate_appendix}),
  we assumed that $\int\varphi^{*}_{\text{TF}}(\vec{r}, t) \hat{p}_z \varphi_{\text{TF}}(\vec{r}, t) d\vec{r} $
  vanishes,
 i.e., that the spatial average of $\hat{H}_1$ vanishes.
 Neglecting the effect of the $\hat{p}_z$ term is consistent with our earlier assumption that the components do not move
 relative to each other during the ramp.
 In addition, we 
  dropped the time-dependent scalar $\int\varphi^{*}_{\text{TF}}(\vec{r}, t)\hat{H}_0\varphi_{\text{TF}}(\vec{r}, t)d\vec{r}$.
  This time-dependent scalar can be rotated away by introducing an overall time-dependent
  phase, i.e., this part of the Hamiltonian does not 
  impact the physics.
    We reiterate that the Hamiltonian $\hat{H}_{\text{TS,t}}$ should work best for fast ramps.

The reduction of the 
GP
description of the lattice system
to the two-state Hamiltonian $\hat{H}_{\text{TS,t}}$  establishes, as discussed in the main text, a connection
between the recoil energy $E_L$, the beginning and end time $\tau$ of the ramp
(assuming a full ramp), and the ramp rate $\alpha$:
$\alpha \tau = 4 E_L$. Thus, for a fixed lattice geometry, a smaller ramp rate $\alpha$ is 
necessarily accompanied by a larger $\tau$. 
Since a larger $\tau$ implies a larger decrease of the mean-field energy during the ramp, one might ask if there are 
alternative approaches to adjusting the ramp rate while maintaining a sufficiently large and approximately
constant mean-field energy during the ramp. 
We now discuss two such approaches.
(i) We performed a sequence of experiments in which the external harmonic trapping potential
was kept on during the ramp. In this case, the BEC does not expand during the ramp and the mean-field energy is maintained. We found that this alternative approach leads to a fair bit of heating during the ramp, in addition to competing dynamics that are 
influenced by the 
harmonic confinement. We concluded that this approach does create more challenges than it solves.
(ii) One could repeat the experiments discussed in our work for lattices with a different 
recoil energy. For example, using a lattice with a larger recoil energy $E_L$ 
would, for fixed $\tau$, translate to a larger $\alpha$ and thus to a smaller dimensionless ramp rate $\tilde{\alpha}$.
The experimental implementation of this is beyond the scope of the present work.

As mentioned in the main text, our 
GP
simulations prepare the initial state $\Psi(\vec{r},-\tau)$ assuming an
axially symmetric trap. Even though the dynamics after turning the external harmonic confinement off, i.e., during
the ramp and subsequent time-of-flight expansion, could---in principle---introduce excitations along the azimuthal angle, this degree of freedom is not treated explicitly in our 
GP
numerics.
The
restriction to an axially symmetric mean-field 
orbital $\Psi(\vec{r},t)$ 
appears to provide
a realistic description of the dynamics.

  \subsection{Considerations related to Fig.~2 of the main text}
  
\begin{figure}
\vspace{0.1cm}
\includegraphics[width=0.45\textwidth]{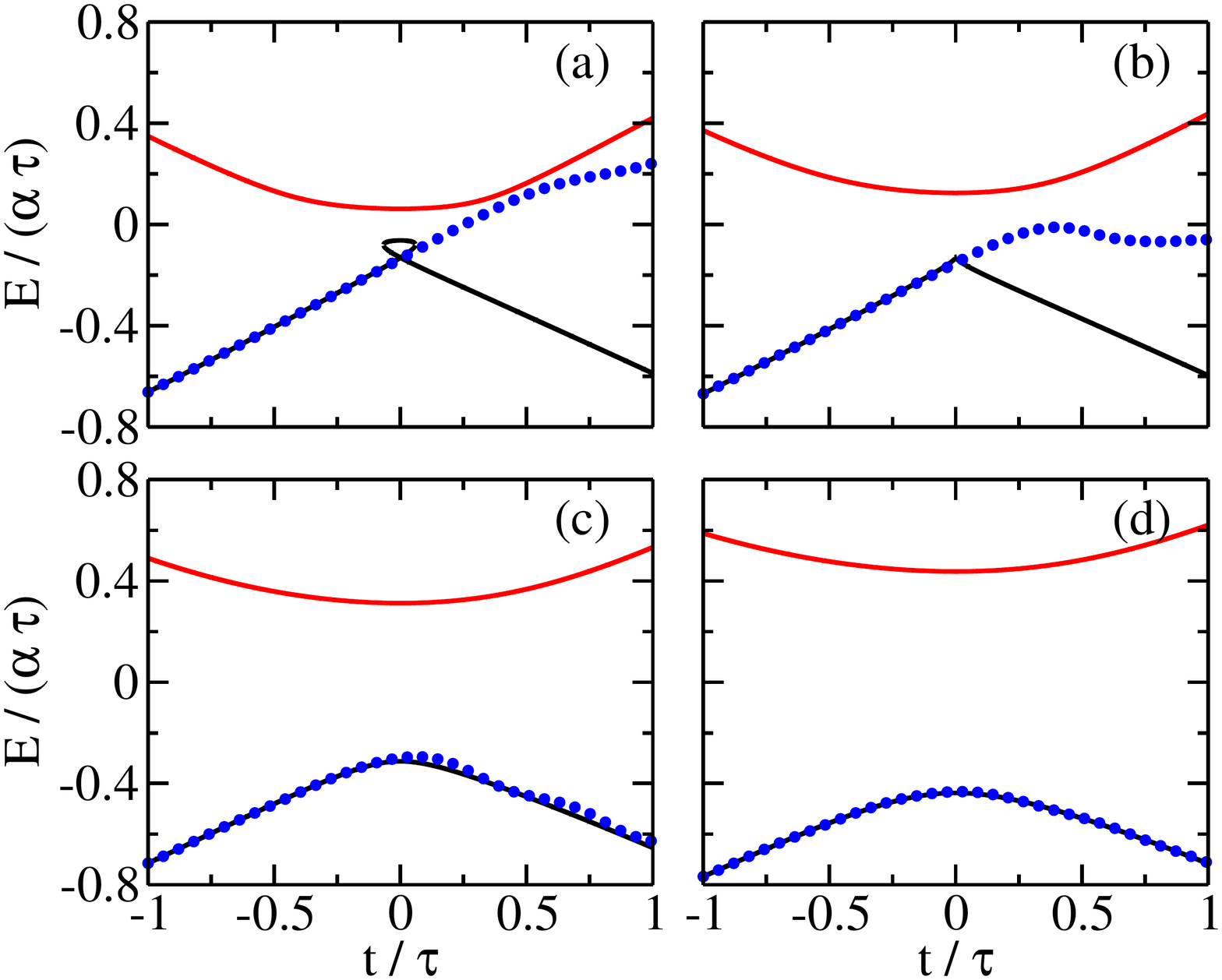}
\caption{Adiabatic and dynamic energy levels
as a function of $t/\tau$
 for
(a) $\tilde{C}(0)/\Omega = 2.14$,
(b) $1.07$,
(c) $0.428$,
and
(d) $0.306$;
the initial states and parameter combinations considered in this figure
are the same as in Fig.~2 of the main text.
The black and red solid lines 
show the lower and upper adiabatic energy levels of $\hat{H}_{\text{TS,t}}$.
The blue circles show the lower dynamic energy level
of $\hat{H}_{\text{TS,t}}$
for 
$\hbar/(\alpha \tau^2) = 7.68 \times 10^{-2}$.
}
\label{fig_sm1}
\end{figure}

  Figure~2 of the main text analyzes the population imbalance
  $\Delta b(t)$ for four different 
  $\tilde{C}(0)/\Omega$ values that range from the mean-field-energy-dominated to the
  lattice-coupling-strength-dominated regime.
  To complement the discussion, Fig.~\ref{fig_sm1}
  shows the adiabatic and dynamic energy levels
  of the two-state Hamiltonian $\hat{H}_{\text{TS,t}}$
  for the same 
  parameters as considered in Fig.~2 of the main text.
  It can be seen that the adiabatic and dynamic energy levels supported by 
  $\hat{H}_{\text{TS,t}}$ 
  agree quite well with those shown in Fig.~1 of the main text.
  Recall, 
  Fig.~1 of the main text uses a constant $C$ value, namely
  $C= \tilde{C}(0)$, but otherwise the same parameters as Fig.~\ref{fig_sm1}.
   As a consequence, the energy gap at 
   $t=0$ in 
  Fig.~\ref{fig_sm1} is equal to that in Fig.~1 of the main text.
  The time dependence of the mean-field energy introduces a slight asymmetry into
  the adiabatic energy levels, i.e., 
  $t=0$ 
  defines no longer
  a symmetry point.
Specifically, compared to Fig.~1 of the main text, 
  the adiabatic energy levels supported by $\hat{H}_{\text{TS,t}}$ are shifted upward
  for 
   $t>0$; the upward shift increases with increasing time
  due to the decrease of $\tilde{C}(t)$.  
    
  Generalizing the derivation presented in Sec.~\ref{two_state}
to  include the $\hbar k_z = 4\hbar k_L$ component,
  we derived a three-state model.  
  The three-state model yields similar results as the two-state model 
  for the parameter regimes considered in this work. 
  This shows that the impact of the higher momentum components is negligible and that the two-state model captures
  the key aspects of the lattice system in the parameter regime considered in this work. 
    
 \subsection{Tunneling probability for the two-state Hamiltonian $\hat{H}_{\text{TS}}$}
 
 The tunneling probabilities for the two-state Hamiltonian $\hat{H}_{\text{TS}}$ were derived in Ref.~\cite{Sniu2}.
 For $C/\Omega \ge 1$, 
 the tunneling probability $r$ is determined by the equation~\cite{Sniu2}
  \begin{eqnarray}
 \label{non_exp}
 \frac{1}{1-r} = \frac{1}{1-\exp\left(-\frac{\pi\Omega^2}{2\hbar\alpha}\right)} + \frac{\sqrt{2}}{\pi}\frac{C}{\Omega}\sqrt{1-r}.
 \end{eqnarray}
 For $C/\Omega <1$,
 one obtains 
 in the adiabatic limit
 the modified exponential Landau-Zener tunneling formula~\cite{Sniu2}
 \begin{eqnarray}
 \label{exp}
 r = \exp\left(-q \frac{\pi\Omega^2}{2\hbar\alpha}\right),
 \end{eqnarray}
 where the scaling factor $q$ is given by
 \begin{eqnarray}
 q=\frac{4}{\pi} \times \nonumber \\
  \int_0^{\sqrt{\left(\frac{\Omega}{C}\right)^{2/3}-1}}(1+x^2)^{1/4}\left[\frac{1}{(1+x^2)^{{3/2}}}-\frac{C}{\Omega}\right]^{3/2}dx.
 \end{eqnarray}
The 
grey-shaded region 
in Fig. 3 of the main text shows Eq.~(\ref{non_exp}) 
with $C$ values ranging from
$C=\tilde{C}(-\tau)$ to $C=\tilde{C}(0)$.
While one might naively think that it would be more appropriate to use 
$C=\tilde{C}(0)$, i.e., the value of the mean-field energy at the mid-point of the ramp, it should be kept in mind that
 the tunneling probability is an integrated quantity
 whose value accumulates during the ramp. Moreover, in the non-linear 
 Landau-Zener model, the tunneling probability at each time depends on the populations and thus indirectly
 on the amount of tunneling that occurred during the earlier part of the ramp. For these reasons, there exists no
 clear argument for how to choose the value of $C$ when using the non-linear two-state 
 model with time-independent $C$ to describe the 
 results obtained using our optical lattice implementation, which is characterized by a time-dependent 
 mean-field energy.

\begin{figure}
\includegraphics[width=0.75\textwidth]{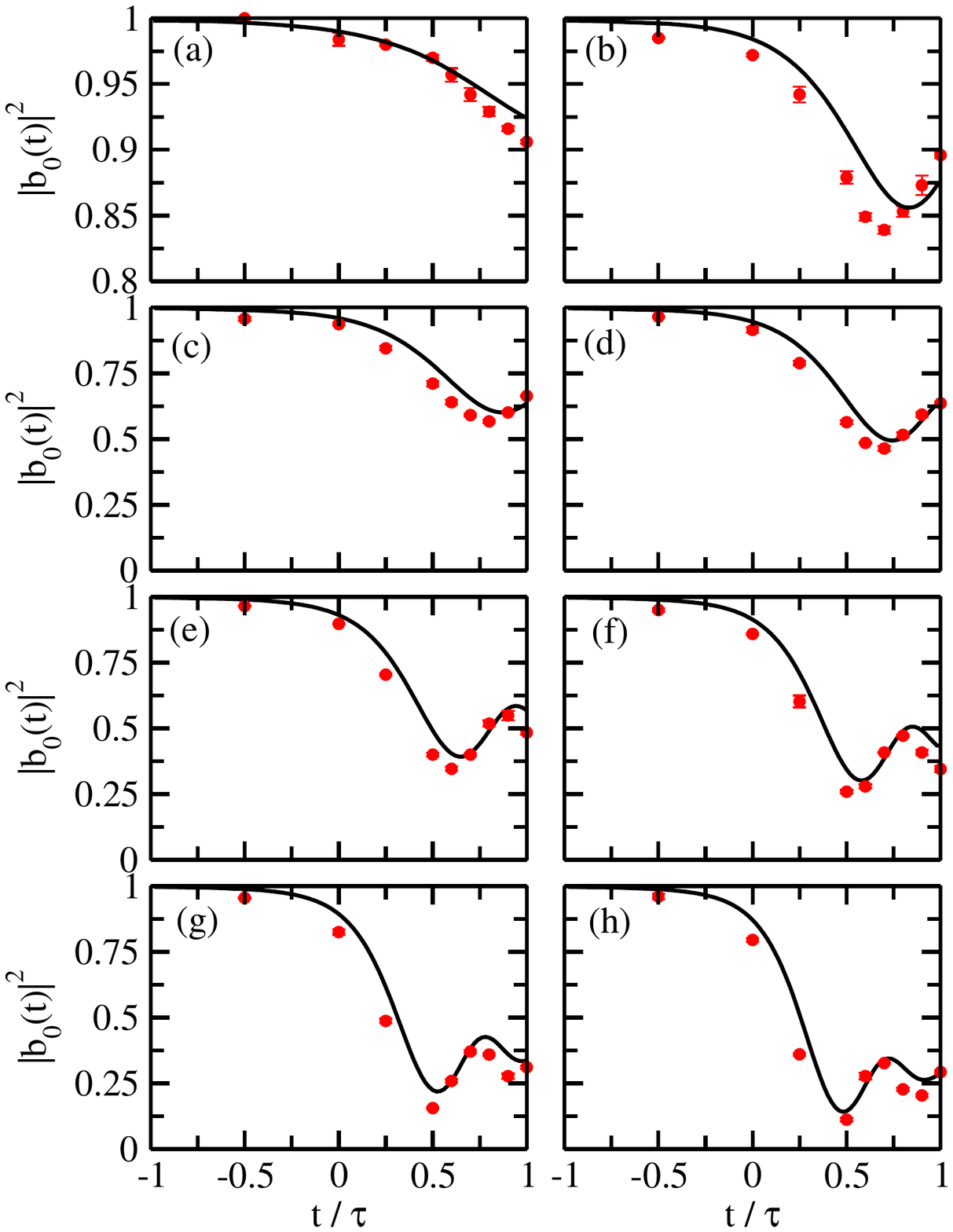}
\vspace*{-0.3in}
\caption{
Extended experimental data sets corresponding to Fig.~3 of the main text;
panels (a)-(h) correspond to the eight (from left to right) experimental data points shown in Fig.~3 of the main text.
Each red circle is the average of three independent experimental runs;
the error bars, which are hardly visible, show the standard deviation.
The uncertainties (not shown) of the populations extracted from the 
GP
simulations
(solid lines) due to the use of finite 
spatial grid spacings and a finite time step are estimated to be smaller than
the experimental error bars.
Note the different ranges of the vertical axis in (a)-(b) and (c)-(h).
\label{fig_sm7}
}
\end{figure}    

\subsection{Extended data sets in the mean-field-dominated regime}

Figure~\ref{fig_sm7} shows extended experimental data sets associated with Fig.~3 of the main text. 
The red circles in Figs.~\ref{fig_sm7}(a)-\ref{fig_sm7}(h) show the experimentally measured
populations $|b_0(t)|^2$  of the zero-momentum component as a function of time for different ramp rates $\alpha$.
It can be seen that the agreement between the 
GP
equation based
results (black lines) and the experimental data (red circles) is quite good.
The data in Fig.~\ref{fig_sm7} show a delayed onset of population
transfer consistent with self-trapping.
The population for the largest time, i.e., for $t=\tau$, is used to make Fig.~3 of the main text.

\subsection{Tunneling probability in the lattice-coupling-strength-dominated regime}

\begin{figure}
\includegraphics[width=0.45\textwidth]{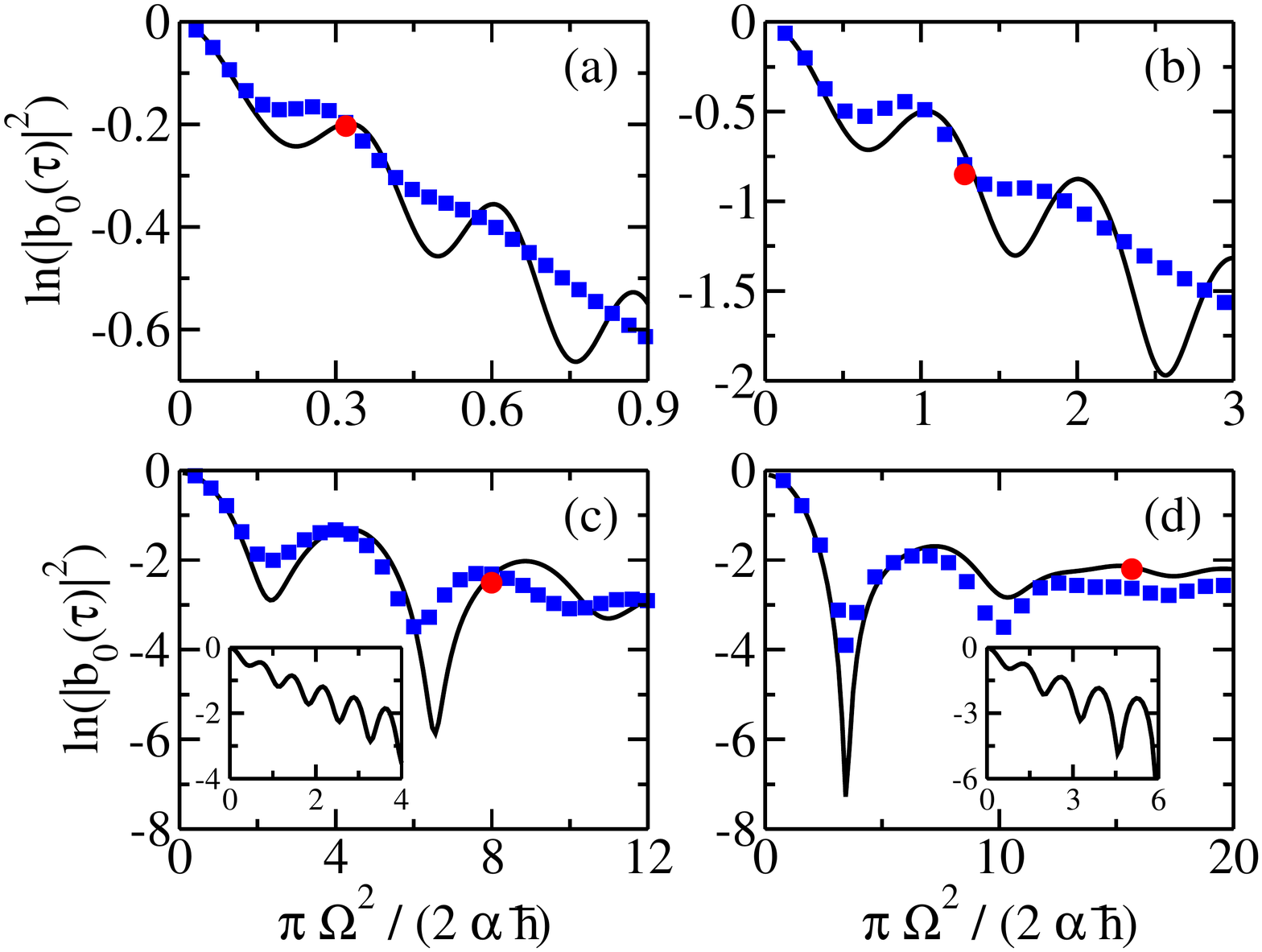}
\caption{Logarithm of $|b_0(\tau)|^2$ as a function of
$\pi \Omega^2 / (2 \hbar \alpha )$
for
(a) $\tilde{C}(0)/\Omega = 2.14$,
(b) $1.07$,
(c) $0.428$,
and
(d) $0.306$;
the initial states and parameters considered in this figure
are, except for $\alpha$, the same as in Fig.~2 of the main text.
The values of $\Omega/(\alpha\tau)$ for panels (a)-(d) are
0.125, 0.25, 0.625, and 0.875, respectively. 
The red circles, which are extracted from Fig.~2 of the main text,
show experimental results  for $\alpha=h\times 9$~$\text{kHz/ms}$ (corresponding to
$\alpha \tau = 4$~$E_L$).
The blue squares show 
GP
results for varying $\alpha$.
For comparison,
the black solid lines show results extracted from the lower dynamic energy level
supported by $\hat{H}_{\text{TS,t}}$.
The insets of (c) and (d) show results extracted from the lower dynamic 
energy level of $\hat{H}_{\text{TS,t}}$
with $\alpha\tau = 12$~$E_L$. 
Note that the ranges of the axis in the insets are different than those in the main figures.
}
\label{fig_sm3}
\end{figure}

Figure~3 of the main text analyzes the tunneling probabilities in 
the mean-field-dominated regime but not in the lattice-coupling-strength-dominated regime.
To elucidate why the 
analysis of the tunneling probabilities in the 
lattice-coupling-strength-dominated regime considered in Figs.~2(c)
[$\Omega/(\alpha \tau) =0.625$]  
and 2(d) 
[$\Omega/(\alpha \tau) =0.875$]  
of the main text is challenging,
Fig.~\ref{fig_sm3} 
plots the logarithm of $|b_0(\tau)|^2$
as a function of $\pi \Omega^2/ (2 \hbar \alpha )$.
The red circles show the experimental data points corresponding
to the linear ramp considered in Fig.~2 of the main text.
The blue squares show results from the 
GP
simulations for different ramp
rates
$\alpha$ but otherwise identical parameters. 
For comparison, 
the black solid lines show the tunneling probability obtained 
using the two-state Hamiltonian $\hat{H}_{\text{TS,t}}$. 
  It can be seen that the tunneling probabilities 
  in Figs.~\ref{fig_sm3}(c) and
  \ref{fig_sm3}(d)
oscillate wildly; 
in particular, the probabilities do not seem to be oscillating around a 
monotonically decaying ``background curve".
The ``wild oscillations" are attributed to the finite time window.
The oscillations become more regular when $\tau$ is, for fixed ramping rate $\alpha$,  increased.
This is illustrated by the
black solid lines in the insets of Figs.~\ref{fig_sm3}(c) and \ref{fig_sm3}(d), 
which show the results for Hamiltonian $\hat{H}_{\text{TS,t}}$
using  $\tau= 12$~$E_L/\alpha $ (i.e., a three times larger $\tau$)
but otherwise identical parameters. 
The oscillation amplitude decreases with increasing $\tau$
while the oscillation frequency increases.
The insets in Figs.~\ref{fig_sm3}(c) and \ref{fig_sm3}(d)
show that $\ln(|b_0(\tau)|^2)$ oscillates around a straight line 
with a slope close to $-1$.
This implies that the tunneling probability 
is, on average and provided $\tau$ is sufficiently large, reasonably well described by the standard linear Landau-Zener formula
in the strong lattice-coupling-strength regime. 
 
\subsection{Integrated densities at the end of the ramp}

The analysis in this section is based on the
GP
orbital
$\Psi(\vec{r},t)$.
Figures~4(a) and 4(b) of the main text show density cuts at the end of the ramp,
i.e., prior to the time-of-flight expansion (recall that 
the ramp sequence is not part of the time-of-flight expansion in our convention), 
while  Figs.~4(c) and 4(d) of the main text show 
integrated densities after the time-of-flight expansion.
Since the ramp time is relatively short, the
two momentum components $\psi_j(\vec{r}, t)$
($j=0$ and $2$) overlap to a good approximation in real space at the end of the ramp. 
To ``isolate" the two components $\psi_j(\vec{r}, t)$, we transform $\Psi(\vec{r},t)$ to momentum space, 
\begin{eqnarray}
\label{momentum_space}
\Phi(\vec{k}, t) = \frac{1}{(2\pi)^{3/2}} \int \Psi(\vec{r}, t)\exp\left(-\imath \vec{k}\cdot\vec{r}\right) d\vec{r}.
\end{eqnarray}
Since $\Phi(\vec{k}, t)$ has two distinct peaks centered around $k_z =0$ and $k_z = 2k_L$,
we define the Fourier-transform of $\psi_0(\vec{r},t)$
as the part of $\Phi(\vec{k}, t)$ with $k_z<k_L$
and the Fourier-transform of $\psi_2(\vec{r},t)$
as the part of $\Phi(\vec{k}, t)$ with $k_z>k_L$,
\begin{eqnarray}
\label{psi_0}
\psi_0(\vec{r}, t) = \frac{1}{(2\pi)^{3/2}}\int \Phi(\vec{k}, t)\Theta(k_L - k_z) d\vec{k}
\end{eqnarray}
and
\begin{eqnarray}
\label{psi_2}
\psi_2(\vec{r}, t) = \frac{1}{(2\pi)^{3/2}}\int \Phi(\vec{k}, t) \Theta(k_z- k_L)d\vec{k},
\end{eqnarray}
where the step function $\Theta(x)$ is equal to $1$ for $x>0$ and equal to $0$ for $x<0$. 

\begin{figure}
 \vspace{1cm}
\includegraphics[width=0.4\textwidth]{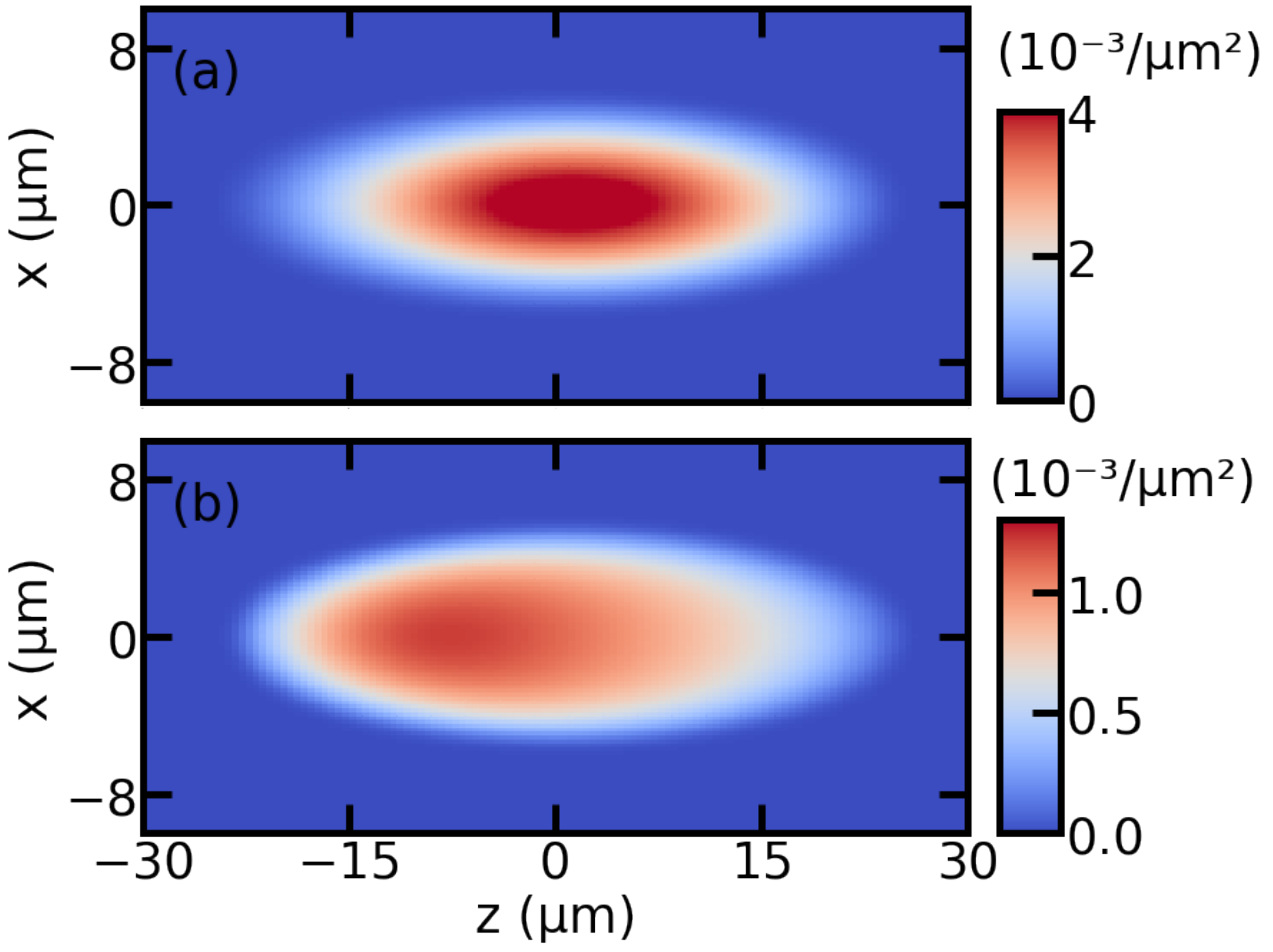}
\caption{Integrated 
GP
densities $n_j(x,z,t)$,
$n_j(x, z, t)=\int_{-\infty}^{\infty} |\psi_j(\vec{r}, t)|^2dy$,
for $t = -\tau + 0.6$~$\text{ms}$, i.e., before the time-of-flight expansion, for 
(a) $j=0$ and (b) $j=2$.
The parameters are the same as those considered in Fig.~4 of the main text.
}
\label{fig_density_sup}
\end{figure}

The cuts shown in Figs. 4(a) and 4(b) of the main text are calculated using Eqs.~\eqref{momentum_space}-\eqref{psi_2}.
To complement the density cuts
discussed in the main text, Figs.~\ref{fig_density_sup}(a) and \ref{fig_density_sup}(b)
show the corresponding integrated densities $n_j(x,z,t)$
for $j=0$ and $j=2$, respectively, for $t=-\tau + 0.6$~ms, i.e.,
prior to the time-of-flight expansion (same
time as the density cuts).
The $n_j(x,z,t)$ are defined by 
integrating over the $y$-coordinate, $n_j(x, z, t)=\int_{-\infty}^{\infty} |\psi_j(\vec{r}, t)|^2dy$. 
The integrated $j=0$ component density is approximately elliptical with a peak located at $\vec{r}=0$.
The maximum of the integrated $j=2$ component density, in contrast, is located at negative $z$.
 The asymmetry of the $j=2$ component density develops during the
 $0.6$~ms short linear ramp, which allows for a tiny movement of the two
 momentum components relative to each other.
 The fact that the internal structures of the $j=0$ and $j=2$ densities 
 differ can be understood by combining the mean-field terms in
 Eqs.~(\ref{coupled_GP_5}) and (\ref{H_0}). This yields
  \begin{widetext}
 \begin{eqnarray}
  (N-1) \begin{pmatrix}
 g |\psi_0(\vec{r},t)|^2 + 2 g  |\psi_2(\vec{r},t)|^2 & 0 \\
 0 & 2 g |\psi_0(\vec{r},t)|^2 +  g  |\psi_2(\vec{r},t)|^2  
 \end{pmatrix};
 \end{eqnarray}
 \end{widetext}
 the ``factor of 2" is due to exchange interactions~\cite{Sguan2020}. For $t$ not much larger than
 $-\tau$, 
 $|\psi_2(\vec{r},t)|^2$ 
 is close to zero. This implies that the dynamics of the $j=0$ component is
 dominated by self-interactions while that of the $j=2$ component is governed by 
 (factor of 2 larger) exchange
 interactions. As a consequence, the $j=2$ component feels an enhanced repulsive mean-field 
 potential
 that is created by the $j=0$ component, thereby explaining the density deformation visible in Fig.~\ref{fig_density_sup}.

 During the time-of-flight expansion, the integrated densities of the two components
 change significantly. 
 This is highlighted by comparing 
 Fig.~\ref{fig_density_sup}(a) and the left parts of Figs.~4(c) and 4(d) of the main text
 as well as
 Fig.~\ref{fig_density_sup}(b) and the right parts of Figs.~4(c) and 4(d) of the main text.
 The dynamics during the time-of-flight
 expansion 
 are influenced by collisions between atoms within each of the two components as well as
 by
 collisions between particles with different $z$-momenta.
 During the expansion, the $j=2$ component ``moves through" the $j=0$ component.
 The dynamics 
 are
 governed by the fact that there exists a tendency to reduce the overlap between the $j=0$ and $j=2$ 
 components. As discussed above, the peak of the $j=2$ density is displaced a bit to 
 negative $z$ at the beginning of the time-of-flight expansion. This imbalance is enhanced during 
 the initial stage of the expansion: with increasing expansion time, the density of the $j=2$ component 
 on the negative $z$ side increases. Eventually, the density becomes too large, creating an energy penalty
 as opposed to an energy reduction.
 As a result, the $j=2$ density redistributes such that the highest density accumulates at the 
 most positive $z$. The above discussion shows that the interplay of the two different momentum space components
 during the ramp and the time-of-flight expansion is due to
 intricate momentum space interactions.

\end{document}